\pdfoutput=1 
\RequirePackage{ifpdf}
\documentclass[notoc,10pt]{JINST}
\usepackage{caption}
\usepackage{subcaption}
\usepackage{wrapfig}
\usepackage{url} 

\usepackage{lineno}

\newenvironment{my_item}{
  \begin{itemize}
    \setlength{\itemsep}{0pt}
    \setlength{\parskip}{0pt}
    \setlength{\parsep}{0pt}
  }{\end{itemize}}

\title{Software digitizer for high granular gaseous detector}

\author{Y. Haddad$^a$\thanks{Corresponding author.}~,
  M. Ruan$^a$
  and V. Boudry$^a$\\
  \llap{$^a$}Laboratoire Leprince-Ringuet, \'Ecole polytechnique,\\
  Palaiseau F-91128, France\\
  E-mail: \email{haddad@llr.in2p3.fr}}

\abstract{ 
  
  A sampling calorimeter equipped with gaseous sensor layers with digital readout \cite{Ammosov:2002jq} is near perfect for ``Particle Flow Algorithm'' \cite{BrientPFA,Videau:2002sk} approach, since it is homogeneous over large surfaces, robust, cost efficient, easily segmentable to any readout pad dimension and size and almost insensitive to neutrons. The response of a finely segmented  digital calorimeter is characterized by track efficiency and multiplicity. Monte Carlo (MC) programs such as GEANT4\cite{Agostinelli2003250} simulate with high precision the energy deposited by particles. The sensor and electronic response associated to a pad are calculated in a separate ``digitization'' process. We developed a general method for simulating the pad response, a digitization, reproducing efficiency and multiplicity,   using the spatial information from a simulation done at higher granularity. The digitization method proposed here has been applied to gaseous detectors including Glass Resistive Plate Chambers (GRPC) and MicroMegas. Validating the method on test beam data, experimental observables such as efficiency, multiplicity and mean number of hits at different thresholds have been reproduced with high precision. 

}


\begin{document}

\section{Introduction}\label{sec:introduction}

One of the main challenges for the future international $e^{+}e^{-}$ linear collider (ILC)~\cite{ILCTDR:Detector} consists in the design of ``Particle Flow Approach'' (PFA)~\cite{BrientPFA,Videau:2002sk} optimised detectors. In this approach, every particle in a given event, including the ones in jets, is reconstructed individually by combining the information of each sub-detector at best. It implies a high longitudinal and transversal granularity in calorimeters to reach a precise jet measurement.

A digital hadron calorimeter~\cite{Ammosov:2002jq} using gaseous sensors such as GRPC or MicroMegas with a digital readout (where one bit is used for each electronic channel) can provide fine segmentation ($\sim 1~\rm cm^2$) for affordable electronic cost. Using hydrogen-free gas, it is also insensitive to the diffuse neutron component of hadronic showers, and constitutes therefore a near perfect ``PFA calorimeter''. 

The raw energy estimation in such calorimeter is done by hit counting, mimicking the ``track length'' counting of the standard calorimetry theory~\cite{Fabjan:1982ev,Fabjan:1985ru}. The response to minimum ionizing particles (MIP) can be characterized by two quantities: the efficiency (the probability to fire at least one pad) and the pad multiplicity (the number of fired pads in each sensor). The multiplicity arises from the induced charge sharing between several pads (marginally by electronic cross talk $\sim$ 0.3 \% measured in~\cite{GRPC:Perf2011}), and might constitutes an important systematic bias in such calorimeters. Especially, in the case of high energy due to the high density of secondary particles. A basic simulation of the particle shower (hadronic or electromagnetic) using GEANT4~\cite{Agostinelli2003250} must be supplemented by a realistic sensor response. An additional module, called ``digitization'', should be added to take into account the physical response of the detector.  

The digitization requires several inputs which are derived from induced charge spectrum distribution and the electronic avalanche size. The measurement of these quantities can be achieved by using reconstructed muon tracks produced by accelerator beam or from cosmic rays.

In this paper, after a short reminder of the gaseous amplification principle, a description of GRPC and
MicroMegas is reported as well as results on the multiplicity and efficiency measurements. A comparison between data and digitized MC-data will be shown and commented upon.

\section{Description of the used gaseous detectors }
\label{sec:detector}

\begin{figure}[!h]
  \begin{center}
    \includegraphics[width=0.8\textwidth]{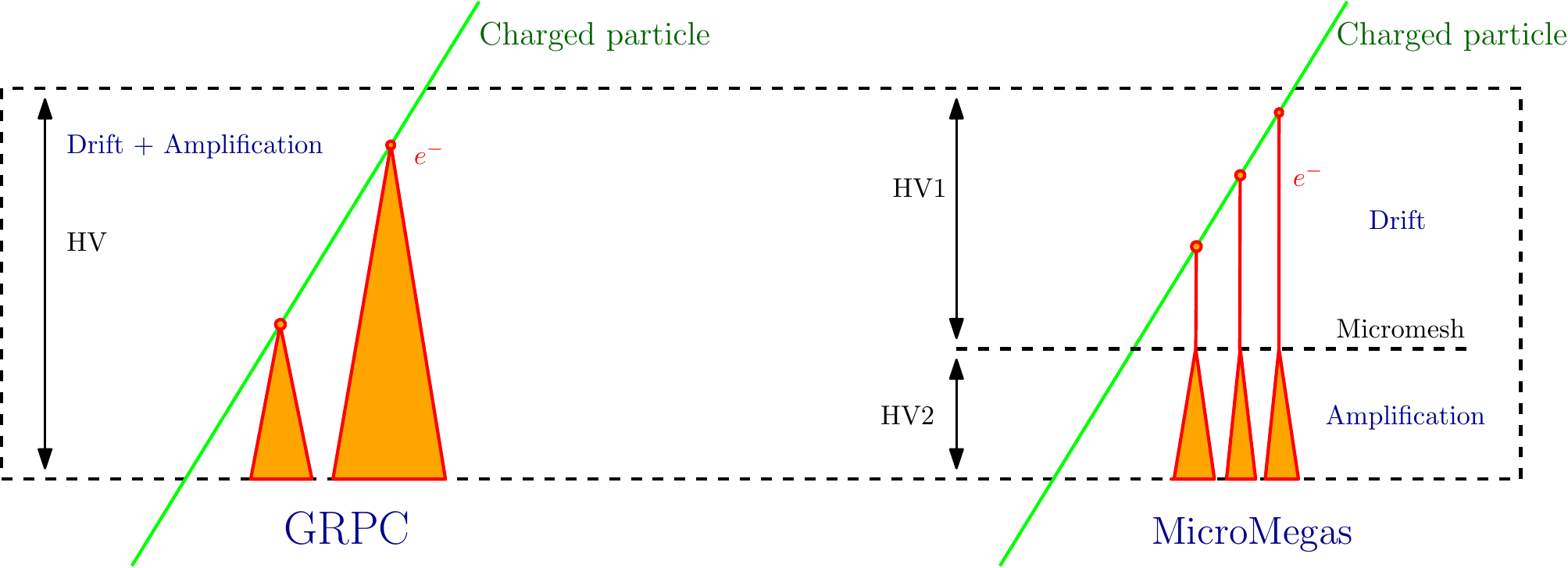}
  \end{center}
  \caption{Schematic of the working principle in the GRPC (left) and MicroMegas (right).}
  \label{fig:WorkingPrinciple}
\end{figure}

Gaseous detectors follow the same working principle. A charged particle passing through the sensor ionizes the gas. Primary electrons drift in the gas gap under an electric field (polarization field) producing a cascade of secondary ionizations, thus a multiplication (amplification) of charge carriers is observed (Figure.~\ref{fig:WorkingPrinciple}). The movement of both electrons and ions leads to an induced charge in the readout pads. In the case of resistive plate chambers, two operating modes can be distinguished, the avalanche and \textit{streamer} modes. In the avalanche mode, the primary ionization leaves a trail of free charge carriers which trigger an avalanche of charges in the electric field (Townsend avalanche). If the size of the avalanche or the polarization field is high enough, a streamer is created. A streamer is the state where photons contribute to the spread of the free charge carriers, a conducting plasma filament can be created between the two electrodes at later stage. Only the avalanche mode is considered as operating mode in GRPC for this study.

Two kinds of sensor have been used here for the digitization study, GRPC and MicroMegas. 
GRPCs consist of two parallel plate electrodes, a positively-charged anode and a negatively-charged cathode polarized by tension of $6.9-7\rm kV$, both made of a very high resistivity glass material (typically $\sim 10^{12}~ \rm \Omega cm$ for glass) and separated by a thin gas volume~\cite{GRPC:Dev2010, GRPC:Perf2011}. The signal is read by pads of $1\times 1~\rm cm^2$ size. A large area of GRPCs ($1~\rm m^2$) with $1.2~\rm mm$ gas gap filled by a typical gas mixture of $93\%$ TFE ($\rm C_2F_4$), $5\%$ $\rm CO_2$ and $2\%$ $\rm SF_6$ operated in avalanche mode provides above $98\%$ of efficiency with pad multiplicity of about $\sim 1.7$ for 0.14~pC threshold~\cite{GRPC:Perf2011}.  

The MicroMegas chambers consist of an anode PCB segmented into $1\times1~\rm cm^2$ pads, the mesh is polarized by a tension of $500~\rm V$~\cite{1748-0221-4-11-P11023}. It uses a gas mixture of $80\%$ $\rm Ar$ and $20\%$ $\rm CO_2$. The amplification and drift gaps are $128~\rm \mu m$ and $3~\rm mm$ respectively. The chambers were tested on in beam condition, as presented in~\cite{1748-0221-5-01-P01013}, and show an efficiency up to $98\%$.

The main differences between GRPC and MicroMegas, stem from the drift and multiplication process. Indeed, in MicroMegas, the drift space is separated from the multiplication area by a polarized mesh. While in the GRPC, the charge carriers drift and multiplication  happen in the same gap, they occur in 2 spaces separated by a polarized mesh in MicroMegas (Figure.~\ref{fig:WorkingPrinciple}). The small multiplication gap of MicroMegas leads also to a better spatial resolution, which implies a better pad multiplicity. 

\subsection{Test beam data}\label{sec:test_beam}
For GRPC the results presented in this paper are based on the data recorded during the beam test campaigns of the SDHCAL~\cite{SDHCALDev,YHaddad:2013} prototype at the SPS CERN facilities in August-September 2012. The experimental setup is composed by 48 active layers and 51 stainless-steel absorbers of 2.8 cm thickness leading for $6$ interaction length. Each sensitive layer is made of a $1\times1 ~\rm m^2$ GRPC, segmented in $1\times1 ~\rm cm^2$ readout pads, for grand total over 450,000 channels for the full detector.

For MicroMegas, the data were taken during the test beam of 2009 in the CERN/PS/T10 zone  as described in~\cite{1748-0221-4-11-P11023}. Four prototypes of $6\times 16\rm ~cm^2$ in addition to one four time larger ($12\times 32 ~\rm cm^2$) active layer were used. 

\section{The determination of induced Charge spectrum}\label{sec:charge_ind}

As mentioned above, the GRPC's were operated in the avalanche mode. Many papers describe the process of avalanche growth and the fluctuation of the induced charge~\cite{GRPC:Lu:2006ht,GRPC:Genz}. Most of those agree with the description of the induced charge spectrum. We chose this description to make comparison with the test beam data. The induced charge spectrum of one MIP in gaseous detector can be estimated from the Polya probability density function (PDF)~\cite{GRPC:Genz} defined by:
\begin{equation}
  {\rm P}(Q_{ind};\theta,\bar{Q}_{ind}) =\left(Q_{ind}\frac{(1+\theta)}{\bar{Q}_{ind}}\right)^{\theta} ~\textrm{exp} \left\{-Q_{ind}\frac{(1+\theta)}{\bar{Q}_{ind}}\right\}
  \label{eq:Polya}
\end{equation}
where $\bar{Q}_{ind}$ and $\theta$ are the mean and the width of the charge function respectively.

In the case of the digital readout, the charge spectrum can be determined by measuring the detection efficiency at different thresholds.

The efficiency of a specific chamber (GRPC or MicroMegas) is defined as the probability to find at least 1 fired pad within 3 cm of a track reconstructed using other chambers \footnote{The track reconstruction is done after merging the hit sharing the common side. The barycenter of formed cluster for a given layer is then considered for track reconstruction \cite{YHaddad:2013}.}. The multiplicity is then the number of matched hits in the chamber.   

The efficiency as the function of threshold $Q_{thr}$ can then be expressed by :
\begin{equation}
  \varepsilon(Q_{thr}) =  \varepsilon_{0} - c \int_{0}^{Q_{thr}}{ {\rm P}(Q_{ind};\theta,\bar{Q}_{ind})} \rm{d}Q_{ind}
  \label{eq:CDFPolya}
\end{equation}
where $\varepsilon_{0}$ is the efficiency of the detector at $Q_{thr} = 0~\rm pC$. A value of $\varepsilon\leq 1$ reflects the limited efficiency of the sensor due to ionization statistic and possible dead channels. 



\begin{figure}[!h]
  \centering
  \begin{subfigure}[b]{0.45\textwidth}
    \centering
    \includegraphics[height=1.2\textwidth]{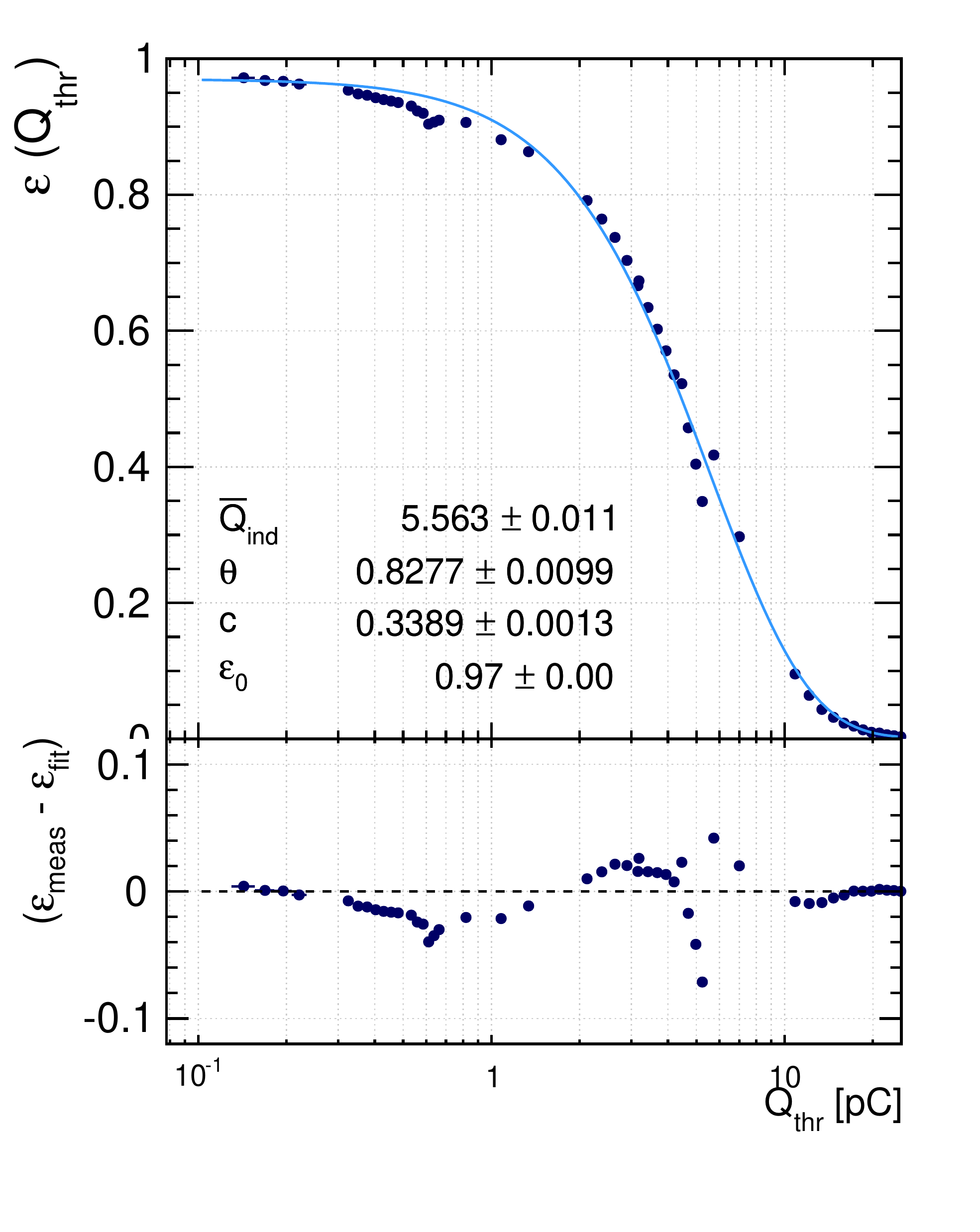}
    \caption{}
    \label{eff_scan_thr}
  \end{subfigure}
  \begin{subfigure}[b]{0.45\textwidth}
    \centering
    \includegraphics[height=1.2\textwidth]{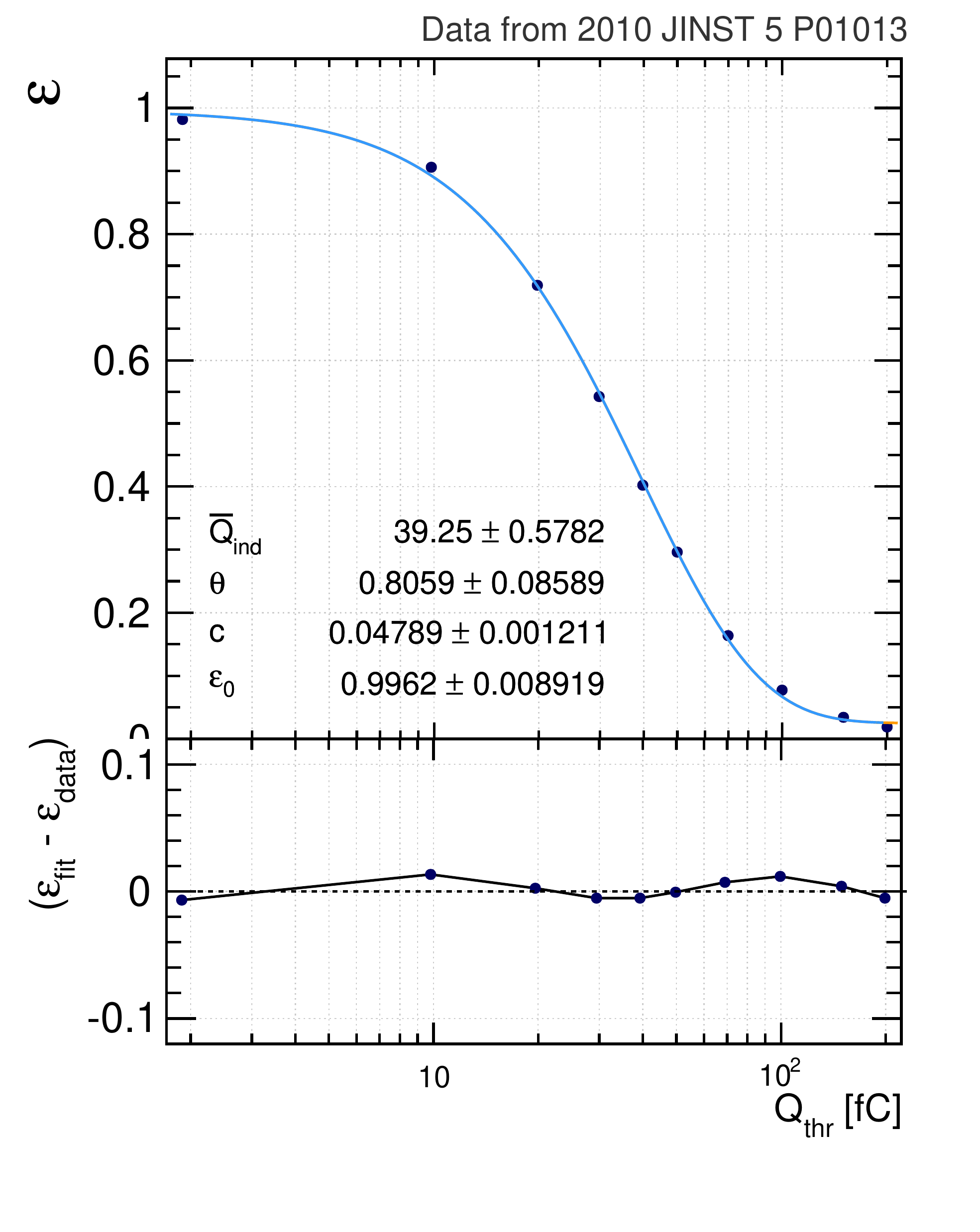}
    \caption{}	    
    \label{eff_mMu}
  \end{subfigure}
  \caption{The GRPC (a) and MicroMegas (b) efficiency versus threshold. The Polya PDF functions are obtained using  equation 3.2. The difference between the data points and the best fit curve in (a) at the value $Q_{thr} = 0.6~\rm pC$ and $Q_{thr}=4-5~\rm pC$ is due to the use of 3 different thresholds regime imperfectly matched.}
  \label{bu}
\end{figure}

The efficiency versus threshold of GRPC measured in 2012 SPS beam test at CERN is shown in Figure \ref{eff_scan_thr}. The efficiency drops from $97\%$ at threshold of $0.1\rm ~ pC$ to less than $1\%$ at high threshold. Using function \ref{eq:CDFPolya} applied in the full range scan, the values of the Polya parameters can be determined. An agreement within $10\%$ between the measured and the fitted efficiency versus threshold is observed. This result validates the choice of a Polya parameterization for the induced charge PDF for GRPC. The mean induced charge $\bar{Q}_{ind}$ in the GRPC is 5.56 pC, with a width of $\theta = 0.82~\rm pC$.  

Concerning MicroMegas, as reported in \cite{1748-0221-5-01-P01013}, the charge spectrum is Landau distributed. However, the Polya parameterization fits the experimental results with an accuracy of less than $3 \%$ (Figure \ref{eff_mMu}). We decided to take Polya PDF as the charge distribution function in MicroMegas. The  mean induced charge in MicroMegas is $39.2 \rm ~fC$, the width is $\theta = 0.8~\rm pC$. 
\subsection{Charge sharing}

\begin{figure}
  \center
  \begin{subfigure}[b]{0.45\textwidth}
    \centering
    \includegraphics[height=0.8\textwidth]{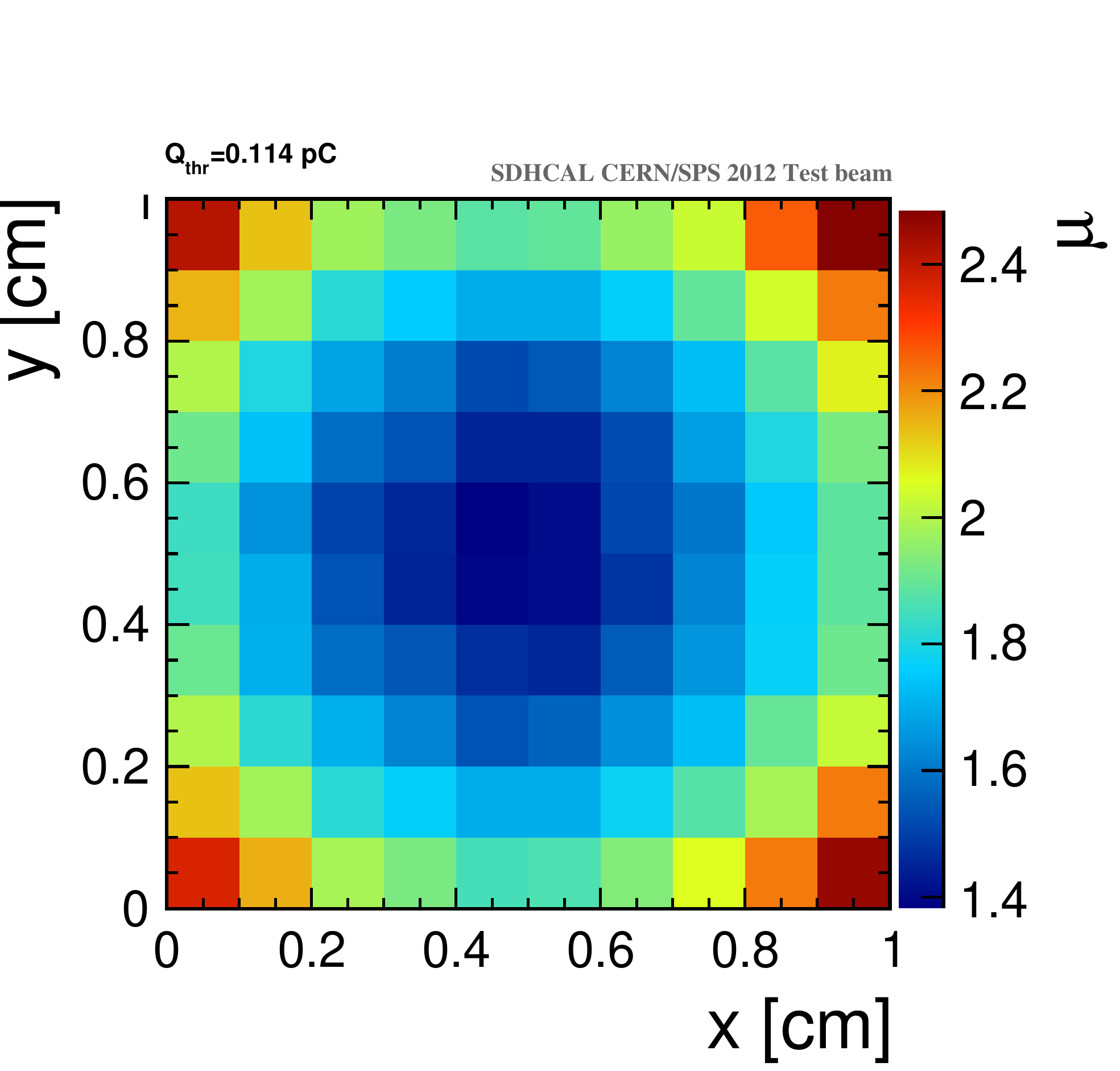}
    \caption{}	    
    \label{pad_scan}
  \end{subfigure}
  \begin{subfigure}[b]{0.45\textwidth}
    \centering
    \includegraphics[height=0.75\textwidth]{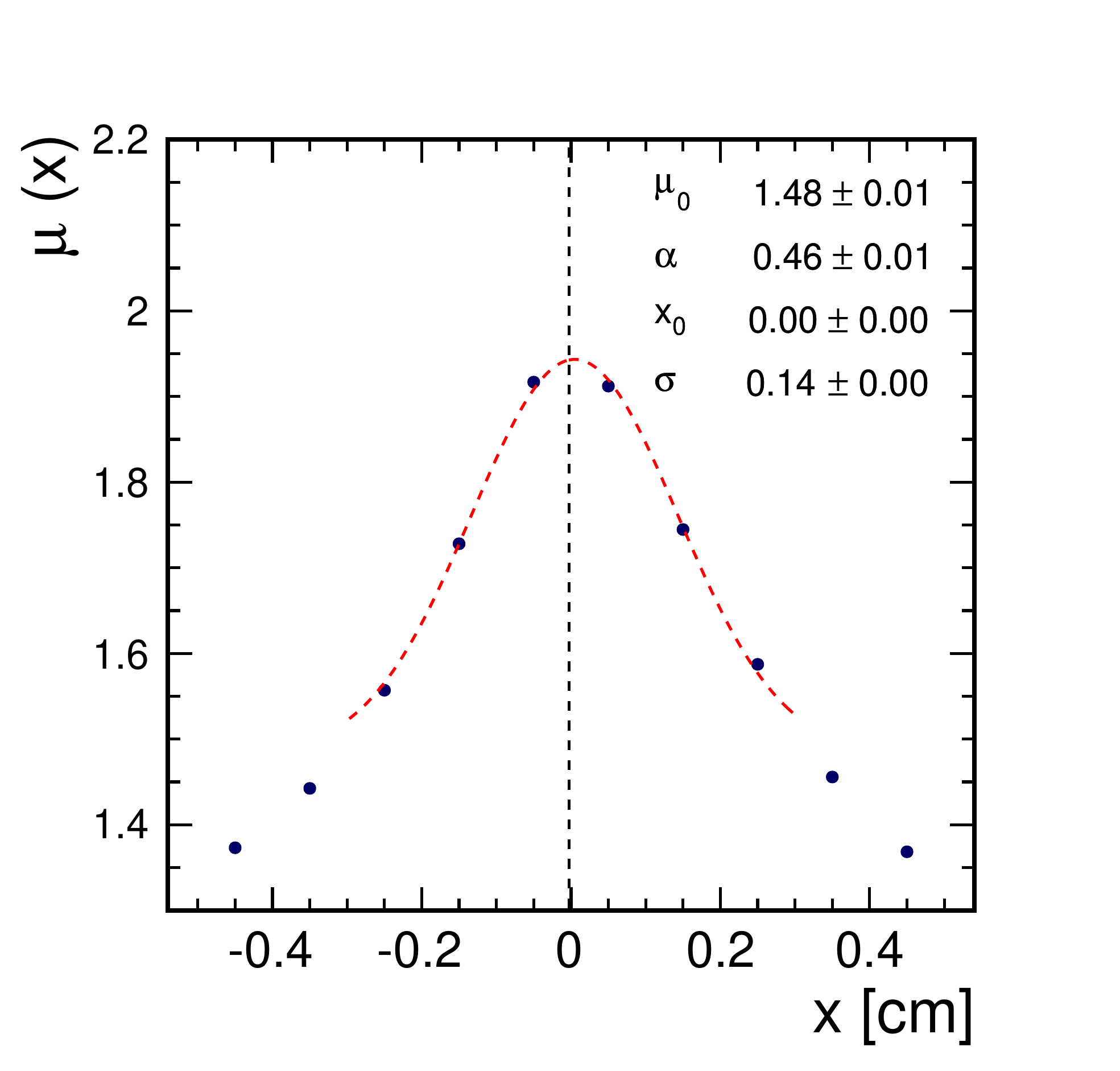}
    \caption{}	    
    \label{2pad_scan}
  \end{subfigure}
  \caption{(\subref{pad_scan}) The multiplicity as a function of the relative MIP position in the pad in GRPC chambers (CERN test beam). (\subref{2pad_scan}) The multiplicity as function of the relative MIP $x$ position at the boundary of two pads, where $0.4<y<0.6~\rm cm$. The red and black lines represent the best Gauss fit  and the pads boundary respectively.}
\end{figure}

The measured average multiplicity of muons, for the GRPC chambers, varies significantly with the position of the reconstructed track in a pad, as can be seen in Figure~\ref{pad_scan}. Multiple pads can be fired if the track position is close to the boundary.


The size of the avalanche can be deduced from the multiplicity distribution: a Gaussian fit \footnote{We use the modified Gauss fit function defined by: $\mu(x|0.4<y<0.6~\rm cm) = \mu_{0}+\frac{\alpha}{\sigma\sqrt{2\pi}}\exp{-\frac{1}{2}\left(\frac{x-x_{0}}{\sigma}\right)^2}$ where $x_{0}$, $\sigma$ and $\alpha$ are the mean, the standard deviation and the normalization of the Gauss function.} on the multiplicity at the center (where $0.4<y<0.6~\rm cm$) of the boundary (Figure. \ref{2pad_scan}) of two adjacent pads shows that the typical size of the charge image is of the order of $\sim 1.4~\rm mm$ (comparable to gas thickness). Thus setting the hit size at 1~mm at simulation level is sufficient to reproduce the multiplicity.



\section{Digitization procedure}\label{sec:digi}

The digitization consists as mentioned previously in reproducing the pad response to given energy deposit (GEANT4 cells). The induced charge has to be determined for each pad independently taking into account the charge sharing between neighbour pads. Starting from the GEANT4 simulation of incoming particle, which provides the position of each interaction in the detector, the procedure can be summarized in a few steps:
\begin{my_item}
\item for each simulated hit, the value of induced charge is drawn randomly following the induced charge spectrum PDF;
\item the charge is distributed to pads following the position of the GEANT4 hit. The fraction of charge attributed to a pad depends on the detector and the pad size;
\item the induced charge above a certain threshold determined for given pad is considered ``fired''.
\end{my_item} 

\begin{figure}[!htbp]
  \centering
  \includegraphics[width=.8\textwidth]{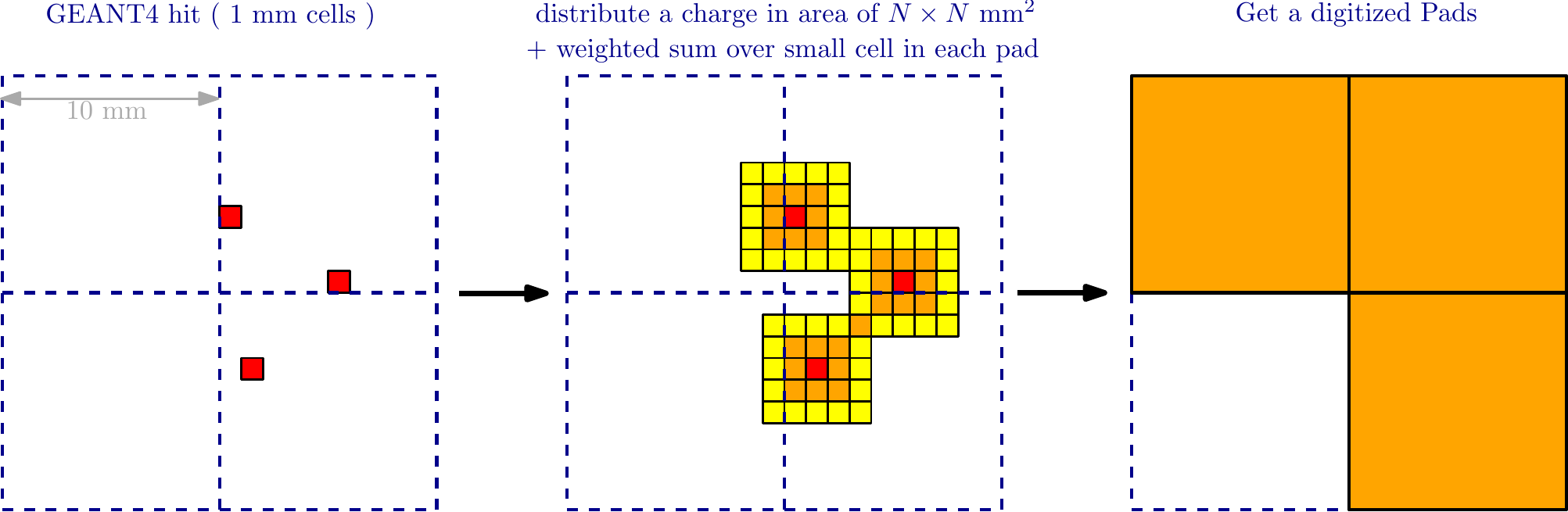}
  \caption{Schematic description of the working principle of the digitizer. A matrix of $5\times 5~\rm mm$ of cell with corresponding weight is created around the simulated cell (GEANT4 hit). A weighted sum is calculated for each pad (bounded by the blue lines). The digitized pads are the ones which reach the threshold.}	    
  \label{small_cell}
\end{figure}

In the approach proposed here, the segmentation of sensitive layer is set at $1~\rm mm$, a GEANT4 hit is regarded as one MIP hit. The chance for having multiple particle hits in the same $1~\rm mm^2$ area is then ignored. The digitization steps are applied to these GEANT4 hits. The fraction of the charge being shared by multiple neighbor pads are calculated from the charge spatial distribution. The surface charge density shape is approximated by a simple 2D Gaussian function. The standard deviation ($\sigma$) of such function is tuned to reproduce the multiplicity and efficiency of each detector. It is summarized into a $N\times N$ sharing fraction table to save computing time. The $N\times N$ weight sharing matrix is determined by an integration of the induced charge density of the avalanche in each cell of $1\times 1 ~\rm mm^2$. The weights are then normalized to the total integral in $N\times N ~\rm mm^2$. The induced charge in each $1\times1~\rm cm^2$ pad is the sum of weights distributed in $N\times N ~\rm mm^2$ sharing matrix around the simulated hit within pad boundaries. When the induced charge in the digitized pad is above a certain threshold, the pad is considered fired. This procedure is summarized by a schematic view in Figure~\ref{small_cell}.

For the GRPC the value of $\sigma $ is set to $\sim 1mm$ (typical lateral induced charge size) and the spatial distribution is carried by $5\times 5 ~\rm mm^2$ weight matrix. 

The spatial distribution of the charge image is parametrized as $3\times3$ matrix for MicroMegas. The center element takes $95\%$ of relative weight. Indeed, the MicroMegas have relatively smaller multiplicity since the amplification region of MicroMegas is much thinner, and directly connected to the readout pads, resulting in a more concentrated charge lateral image. 

\section{Results and discussion}

\begin{figure}[!h]
  \centering
  \begin{subfigure}[b]{0.51\textwidth}
    \centering
    \includegraphics[height=.77\textwidth]{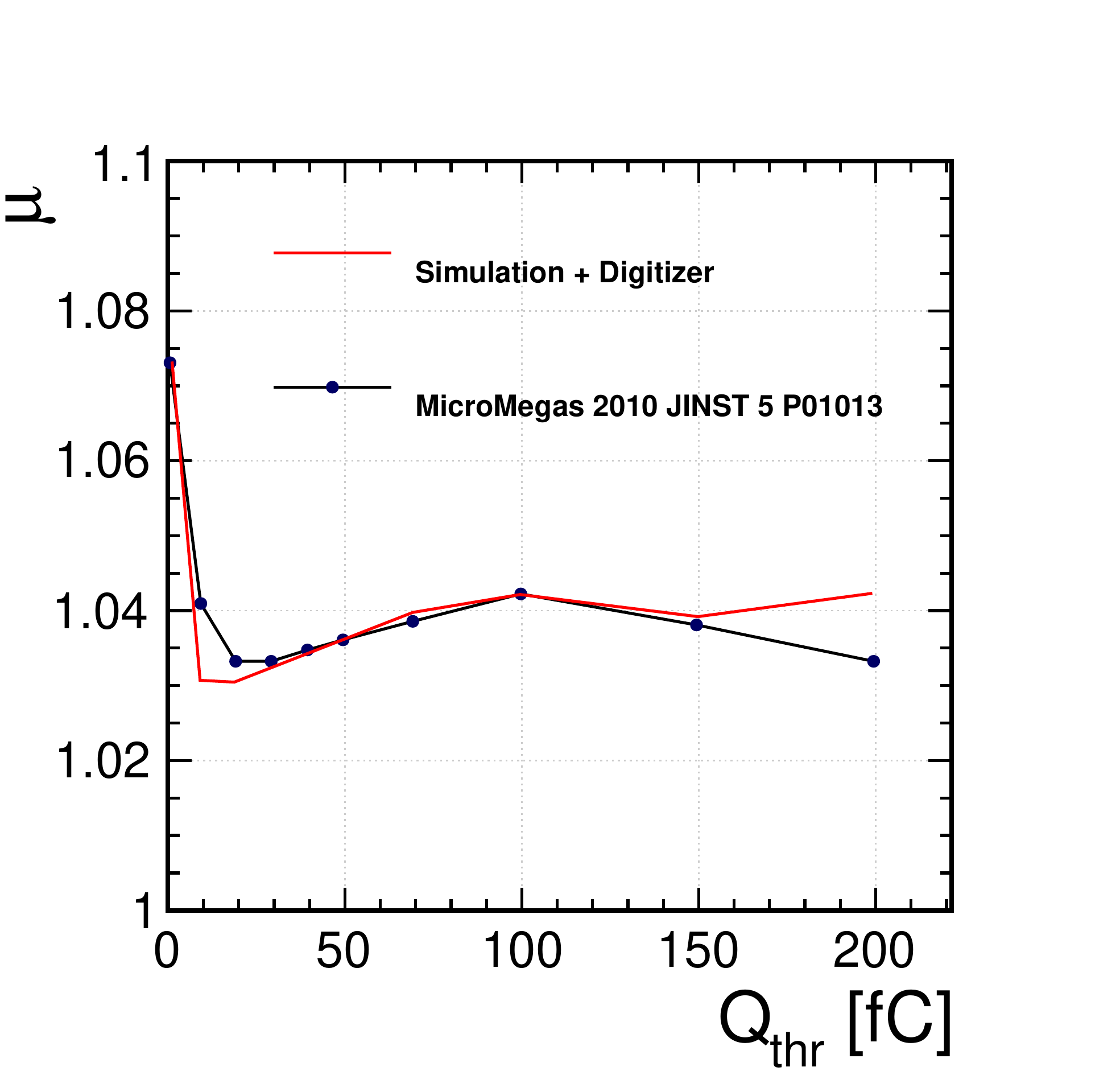}
    \caption{}	    
    \label{mul_muM}
  \end{subfigure}
  \begin{subfigure}[b]{0.45\textwidth}
    \centering
    \includegraphics[height=.88\textwidth]{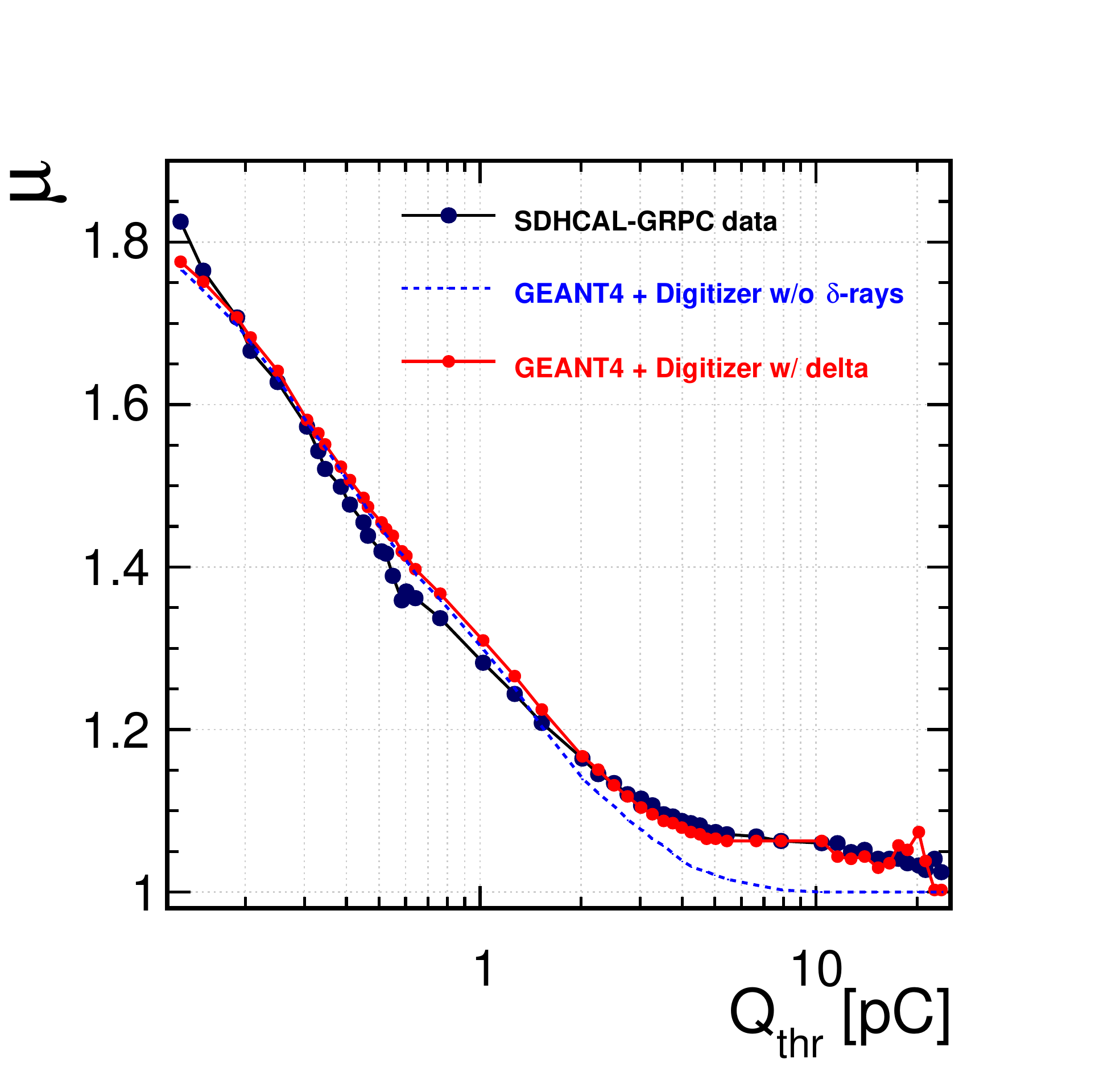}
    \caption{}
    \label{mul_GRPC}
  \end{subfigure}
  \caption{ Multiplicity of MicroMegas (a) and GRPC (b) versus threshold. The red and the black lines represent the simulation ( using GEANT4 within the digitizer) and sensor data. The dashed blue line in (b) represents the use of the digitizer without including the $\delta$-rays effect.}
  \label{bu}
\end{figure}

The validation of the digitization procedure proposed here is tested on MicroMegas and GRPC data. As mentioned above the digitization is the ability to reproduce the pad multiplicity at different threshold values. The scan done in the MicroMegas test beam presented in \cite{1748-0221-4-11-P11023} is well reproduced (Figure~\ref{mul_muM}). The pad multiplicity takes the value $1.07$ at $1~\rm fC$ threshold, and rapidly decreases to a level of $1.03$ at high values. An increasing of about $1\%$ of the multiplicity is observed around $100~\rm fC$ creating a small bump. According to \cite{1748-0221-4-11-P11023}, the bump is induced by knock-on electrons ($\delta$-rays). Indeed, induced charges above $30~\rm fC$ are mainly originated from events with high energy deposit. These likely produce $\delta$-rays leading to some ionization on the neighboring cells and hence a higher multiplicity.




For a charged particle with $\beta\sim 1$, the probability to have a knock-on electron with energy above $1~\rm keV$ in 1mm of Ar gas is about $2\%$~\cite{bichsel2006method}. This is modeled in the digitizer by amplifying the induced charge of the neighbor cells for $2\%$ of cases randomly. The induced charge in the neighbor cell is then the sum of the initial induced charge ($Q_{neighbor}$) and the total average of the induced charge in the sensor ($\bar{Q}_{ind}$) with a standard deviation\footnote{The induced charge of the $\delta$-rays is supposed to be Gauss distributed} of $\sigma_{\delta}$: $Q'_{neighbour}=Q_{neighbour} +( \bar{Q}_{ind}\pm \sigma_\delta)$. The result shown in Figure \ref{mul_muM} is obtained for $\sigma_{\delta}\sim 2.2$. The deviation at high threshold is probably due to the low statistic in this region. 

The same measurement was also made for GRPC sensors, where the multiplicity was determined for each threshold value in the range $[0.1,25] \rm ~pC$ using $30\rm ~GeV$ muons. The pad multiplicity takes the value of $1.82$ at $0.13\rm pC$ and drops drastically to $1.05$ at $Q_{thr}\sim 4~\rm pC$ before reaching a plateau in between $4~\rm pC$ -- $11\rm ~pC$ and decreases again to $1$. The plateau is probably due to the presence of the $\delta$-rays. Following the same approach as MicroMegas the $\delta$-rays effect can be modeled by inducing the same amount of $\bar{Q}_{ind}$ for GRPC in a neighbor cell with $\sigma_\delta\sim 5$, since this value gives the best agreement with the data.

\begin{figure}
  \centering
  \includegraphics[width=.5\textwidth]{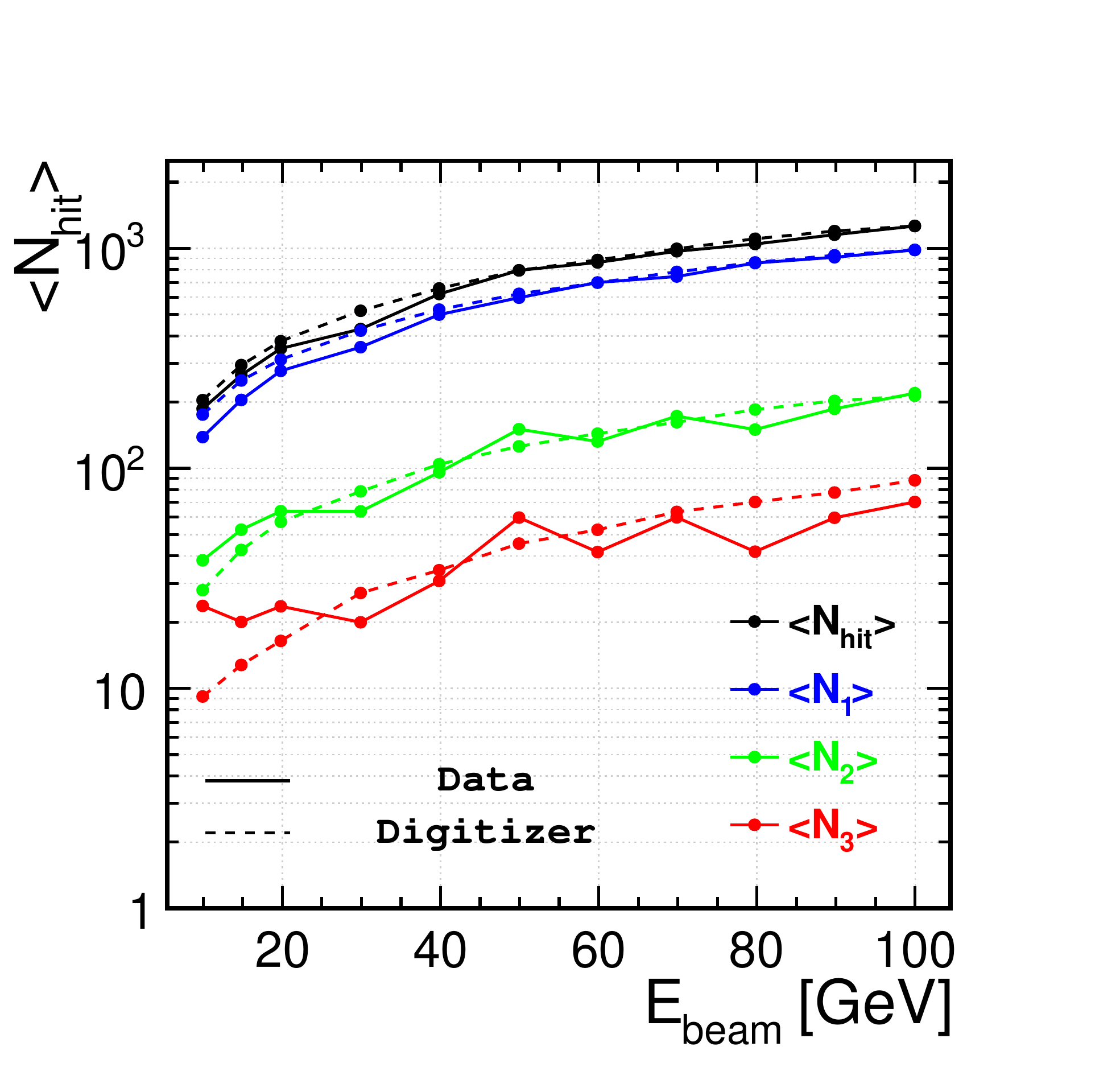}
  \caption{Number of hits versus pion beam energy for SDHCAL prototype. The blue, green and red curves represent the number of hit reaching only the corresponding threshold values 0.114pC, 5pC, 15pC, respectively. The black curves are the sum of all hits. The data correspond to the solid line, the simulation to the dashed one.}
  \label{nhit_thr}
\end{figure}

Samples of pions using GPRC calorimeter prototype are also considered. Simulated samples are produced with the same particle type in the energy range of $10 \rm ~GeV$ up to $100 \rm ~GeV$ according to the test beam data. The digitizer reproduces the pion response (number of calorimeter hits) in the calorimeter as shown in figure \ref{nhit_thr} for three different threshold values ($0.114~\rm pC$ ,$ 5\rm ~pC$, $15\rm ~pC$). For the highest threshold, the data have a large dispersion relative to the predicted value by the digitizer. This behavior is believed to be due to the statistical fluctuations between the 48 chambers at high threshold. The performances achieved here are sufficient for MC-data comparison of complete shower in the detectors.  

The measurement of the pion response was only done for the GRPC data where the experimental setup allows to measure the response of hadron showers \cite{YHaddad:2013}, unlike MicroMegas which have been not tested as calorimeter. 

\section{Conclusion}

The digitization -- or the modelisation of sensors and electronic associated to a readout pad -- is an indispensable and complementary part of generic simulation, such as GEANT4. We developed a general gaseous detector digitizer which can reproduce the efficiency and multiplicity response to a minimum ionizing particle (MIP), two key response parameters for gaseous calorimeter with a digital readout, using a very fine sensor segmentation.

Polya function is used to model the induced charge spectrum using only two parameters. It describes, with good agreement, the evolution of the efficiency versus the threshold, for both thin GRPC and MicroMegas detectors.

The finite spatial distribution of the induced charge by a MIP track results in the firing of neighbouring pads. The expected number of fired pads highly depends on the relative position within crossed pad. Using a pad sub-segmentation of $1~\mathrm{mm}$, and sharing the charge with the proper distribution in conjunction with the charge spectrum, the digitizer reproduces adequately the efficiency and multiplicity on MIPs (muons) and pions data.


\bibliography{mybib}{}
\bibliographystyle{ieeetr}
\end{document}